\documentstyle[seceq,twoside]{ptptex}
\notypesetlogo  
\title{Covariant Level-Classification Scheme and Chiral Symmetry}
\author{%
Shin {\sc Ishida}, Muneyuki {\sc Ishida}$^{*}$ and Tomohito {\sc Maeda} }
\inst{%
 Atomic Energy Research Institute, 
College of Science and Technology\\
Nihon University, Tokyo 101-0062, Japan\\
$^{*}$Department of Physics, Tokyo Institute of Technology\\
Tokyo 152-8551, Japan
}
\abst{%
Starting from the bi-local Klein-Gordon Equation with spin-independent squared-mass
operator, we give a covariant
quark representation of general composite meson systems with definite 
Lorentz transformation properties. For benefit of this representation
we are able to deduce automatically the transformation rules of composite
mesons for general symmetry operations from those of constituent (exciton)
quarks. Applying this we investigate especially physical implication 
of chiral symmetry for the meson systems, and present a covariant level-classification
scheme, leading to a possible existence of new meson multiplets of ``chiralons."
}

\begin{document}
\maketitle

\setcounter{tocdepth}{4}

\section{Introduction}

There are the two contrasting view points of composite quark-antiquark
mesons: The one is non-relativistic, based on the approximate symmetry of
$LS$-coupling in the non-relativistic quark model (NRQM); 
while the other is relativistic, based on the 
dynamically broken chiral symmetry 
typically displayed in the Nambu Jona-Lasinio (NJL) model. 
The $\pi$-meson (or $\pi$-nonet) is now widely believed to have a
dual nature of non-relativistic system 
and also of relativistic system as a Nambu-Goldston (NG) boson 
in the case of spontaneous breaking of chiral symmetry. However,
no successful attempts to unify the above two view points 
have been yet proposed.

On the other hand we have developed the 
covariant oscillator quark model (COQM)\cite{rf1,rf3,rf2}
for many years as a covariant extension of NRQM, which is based on 
the boosted $LS$-coupling scheme. The meson wave functions (WF)
in COQM are the restricted tensors in the $\tilde U(4)\bigotimes O(3,1)$ space
which reduce at the rest frame to those in the 
$SU(2)_{\rm spin}\bigotimes O(3)_{\rm orbit}$ space in NRQM. Although
the COQM had been successfully applied to various non-static as well as static problems
of hadrons with general quark configuration,
no 
consideration on chiral symmetry was given there.

The purpose of this paper is
 to get rid of this defect in COQM and 
is to give a unified view point of the two contrasting ones
of the composite meson systems,
extending the WF to the general tensors 
in the  $\tilde U(4)\bigotimes O(3,1)$ space,
which are required for taking into account chiral symmetry.

\section{Covariant Framework for Describing Composite Mesons}
For meson WF described by $\Phi_A{}^B(x_1,x_2)$ ($x_1,x_2$ denoting the 
space-time coordinate and 
$A=(\alpha ,a)(B=(\beta ,b))$ denoting the Dirac spinor and 
flavor indices  of constituent quark (anti-quark)) 
we set up the bilocal Yukawa equation\cite{rf2}
\begin{eqnarray}
\left[ \frac{\partial^2}{\partial X_\mu^2}
 - {\cal M}^2(x_\mu ,\frac{\partial}{\partial x_\mu} )\right]
\Phi_A{}^B(X,x)=0
\label{eq1}
\end{eqnarray}
($X(x)$ denoting the center of mass (CM) (relative) coordinate of meson),
where the  ${\cal M}^2$ is squared mass operator including
only a central, Dirac-spinor-independent 
confining potential.
The WF is separated into the plane wave describing 
CM motion and the (Fierz-component) internal WF as 
\begin{eqnarray}
\Phi_A{}^B(x_1,x_2)=\sum_{{\bf P}_n,n}(e^{iP_nX}\Psi_{n,A}{}^{(+)B}(x,P_n)
+e^{-iP_nX}\Psi_{n,A}{}^{(-)B}(x,P_n)),
\label{eq2}
\end{eqnarray}
where the Fierz components $\Psi_n^{(\pm )}$ are eigenfunctions
 of ${\cal M}^2$;  
$P_{n,\mu}^2=-M_n^2,\ P_{n,0}=\sqrt{M_n^2+{\bf P}_n^2}$;
and the label 
$(\pm )$ represents the positive (negative) frequency part;
and $n$ does a freedom of excitation. We have the following field 
theoretical expression for the WF in mind 
as a guide for the present semi-phenomenological approach:
\begin{eqnarray}
\Phi_A{}^B(x_1,x_2) &=& \sum_n[\langle 0|\psi_A(x_1)\bar\psi^B(x_2)|M_n\rangle
+\langle M_n^c|\psi_A(x_1)\bar\psi^B(x_2)|0\rangle ] ,
\label{eq3}
\end{eqnarray}
where $\psi_A(\bar\psi^B)$ denotes the quark field (its Pauli-conjugate)
and $|M_n\rangle\ (\langle M_n^c|)$ does the composite meson 
(its charge conjugate) state.
The internal WF is expanded in terms of 
a complete set $\{ W_i \}$ of free bi-Dirac spinors of 
quarks and anti-quarks as
\begin{eqnarray}
\Psi_A^{(\pm )B}(x,P_n)=\sum_{i}W_{i\alpha}^{(\pm )\beta}(P_n)
\phi_a^{(\pm )b}(x,P_n) .
\label{eq4}
\end{eqnarray}

\section{Complete Set of Spin Wave Function 
and Irreducible Meson Components}

We set up the conventional ``free'' Dirac quark-spinors with four-momentum of 
composite meson itself $P=P_M$,
$u_q(P_\mu ,s_q)$, which satisfies the Dirac equation,
$(iP\cdot\gamma + M)u_q(P_\mu ,s_q)=0,$ 
where $s_q$ represents the spin up-down of the quark and 
$P_0 > 0$ is the energy of the meson.
In \S 4 we consider the chiral transformation property of composite meson systems.
Through chiral transformation, $u_q (P_\mu ,s_q)$ is related with  
$\gamma_5\ u_q (P_\mu ,s_q)$, which has the energy-momentum  $-P_\mu$,
since $\gamma_5\ u_q (P_\mu ,s_q)$ satisfies the Dirac equation
$(-iP\cdot\gamma +M)\  \gamma_5\ u_q (P_\mu ,s_q)=0$. Thus, 
$\gamma_5\ u_q (P_\mu ,s_q)\sim u_q (-P_\mu ,s_q)$.
We define $(u_q(P_\mu ,s_q),u_q(-P_\mu ,s_q))$, 
both of which are the solutions of the above Dirac equation, as
$
(u_q(P_\mu ,s_q),u_q(-P_\mu ,s_q)) 
 = (u({\mib P},s_q),-s_qv({\mib P},-s_q)).$

Conventional ``free'' Dirac antiquark-spinors with four-momentum $P_\mu$,
$\bar v^{\bar q}(P_\mu ,s_{\bar q})$, satisfies the Dirac equation,
$
\bar v^{\bar q}(P_\mu ,s_{\bar q})(-iP\cdot\gamma + M)=0,
$ 
where $s_{\bar q}$ represents the spin up-down of the antiquark.
We define $(\bar v^{\bar q}(P_\mu ,s_{\bar q})$, 
$\bar v^{\bar q}(-P_\mu ,s_{\bar q}))$, 
both of which are the solutions of the above Dirac equation, as
$
(\bar v^{\bar q}(P_\mu ,s_{\bar q}),\bar v^{\bar q}(-P_\mu ,s_{\bar q})) 
  =  (\bar v({\mib P} ,s_{\bar q}),s_{\bar q} \bar u({\mib P},-s_{\bar q})).
$

It is to be noted that all four spinors for both ``quarks 
and anti-quarks"  are necessary
 to describe the spin WF of mesons.
Then the complete set of bi-Dirac spinors $W^{(+)}(P)$ is 
given by
\begin{eqnarray}
U(P) &=& u_q(p_1,s_q)\bar v^{\bar q}(p_2,s_{\bar q})|_{p_{i,\mu}=\kappa_iP_\mu}
=u({\mib P},s_q)\bar v({\mib P},s_{\bar q}),\nonumber\\
C(P) &=& u_q(p_1,s_q)\bar v^{\bar q}(-p_2,s_{\bar q})|_{p_{i,\mu}=\kappa_iP_\mu}
=u({\mib P},s_q)\bar u({\mib P},-s_{\bar q})s_{\bar q},\nonumber\\
D(P) &=& u_q(-p_1,s_q)\bar v^{\bar q}(p_2,s_{\bar q})|_{p_{i,\mu}=\kappa_iP_\mu}
=-s_q v({\mib P},-s_q)\bar v({\mib P},s_{\bar q}),\nonumber\\
V(P) &=& u_q(-p_1,s_q)\bar v^{\bar q}(-p_2,s_{\bar q})|_{p_{i,\mu}=\kappa_iP_\mu}
=-s_q v({\mib P},-s_q)\bar u({\mib P},-s_{\bar q}) s_{\bar q},\ \ \ \ \ \ \ \ \ \ 
\label{eq9}
\end{eqnarray}
where
in Eq.(\ref{eq9}) 
we have defined technically the momenta of
``constituent quarks"\footnote{
In so far as concerned with Eqs.~(\ref{eq9}) and (\ref{eq10}) the quantities $\kappa_i$
and accordingly $m_i$ are arbitrary and have no physical meaning. However, $m_i$
have proved to be the effective masses of constituent quarks through the phenomenological
applications of COQM so far made.
} as 
\begin{eqnarray}
p_{i,\mu} &\equiv&  \kappa_iP_\mu ,\ p_{i,\mu}^2=-m_i^2;\ P_\mu^2=-M^2,
M=m_1+m_2\nonumber\\
 (\kappa_{1,2} &\equiv& m_{1,2}/(m_1+m_2); \ \ \kappa_1+\kappa_2=1).
\label{eq10}
\end{eqnarray}
The respective members in Eq.(\ref{eq9}), 
Non-Relativistic, $\bar q$-type and $q$-type Semi-Relativistic, 
and Extremely Relativistic ones, 
are expressed 
in terms of their irreducible composite meson WF 
as follows:
\begin{eqnarray}
  U_A{}^B(P) &=& \frac{1}{2\sqrt{2}}
   [(i\gamma_5P_{s,a}^{(NR)b}(P)+i\gamma_\mu V_{\mu ,a}^{(NR)b}(P))
    (1+\frac{iP\cdot\gamma}{M}) ]_\alpha{}^\beta ,\nonumber \\
  C_A{}^B(P) &=& \frac{1}{2\sqrt{2}}
   [(S_{a}^{(\bar q) b}(P)+i\gamma_5\gamma_\mu A_{\mu ,a}^{(\bar q)b}(P))
    (1-\frac{iP\cdot\gamma}{M}) ]_\alpha{}^\beta ,\nonumber\\
  D_A{}^B(P) &=& \frac{1}{2\sqrt{2}}
   [(S_{a}^{(q) b}(P)+i\gamma_5\gamma_\mu A_{\mu ,a}^{(q)b}(P))
    (1+\frac{iP\cdot\gamma}{M}) ]_\alpha{}^\beta ,\nonumber\\
  V_A{}^B(P) &=& \frac{1}{2\sqrt{2}}
   [(i\gamma_5P_{s,a}^{(ER)b}(P)+i\gamma_\mu V_{\mu ,a}^{(ER)b}(P))
    (1-\frac{iP\gamma}{M}) ]_\alpha{}^\beta ,\ \ \ \ \ \ \ \ \ \ \ 
\label{eq11}
\end{eqnarray}
where  
all vector and axial-vector mesons satisfy the Lorentz 
conditions, $P_\mu V_\mu (P)=P_\mu A_\mu (P)=0$.
Here it is to be noted that, in each type of the above members, 
the number of freedom counted both in the quark representation and 
in the meson representation is equal,
as it should be ($2\times 2=4$ and $1+3=4$, respectively).

\section{Level Classification and Chiral Symmetry}
\subsection{Level structure of ground state mesons}
Thus far we have presented a general covariant kinematical
framework for describing the (ground states of) composite meson systems.
However, what kinds of mesons do really exist or not, that is, the meson
spectroscopy, is a dynamical (still unsolved) problem of QCD.

For this problem, we follow a physical idea of dynamically
broken chiral symmetry of QCD, typically displayed in the NJL model:
In the heavy quarkonium ($Q\bar Q$) system both quarks($Q$) and 
antiquarks($\bar Q$) are possible to do, since $m_Q > \Lambda_{\rm conf}$, 
only non-relativistic motions with positive energy,
and the $LS$-symmetry is good.
Accordingly the bi-spinor $U$ is considered to be applied to $Q\bar Q$
 system as a covariant spin WF.
In the heavy-light quark meson $Q\bar q$($q\bar Q$) system
the $\bar q$($q$) make, since   $m_q \ll \Lambda_{\rm conf}$, 
relativistic motions both with positive and negative energies,
and the chiral symmetry concerning $\bar q$($q$) is good.
Accordingly both the $U$ and $C$ ($U$ and $D$)
are to be applied to the  $Q\bar q$($q\bar Q$) system, leading to 
possible
existence of new composite scalar and axial-vector mesons(see
Eq.(\ref{eq11})).
In this meson system it is to be noted that the conventional heavy
quark symmetry (HQS) is also valid as an approximate symmetry.
In the light quark $q\bar q$-meson system 
both quarks $q$ and anti-quarks $\bar q$
make, since $m_q\ll\Lambda_{\rm conf}$, relativistic motion 
with both positive and negative energies, and chiral symmetry is good.
Accordingly the linear combinations of
$U$ and $V$ are applied to the $q\bar q$-system, and
in this system there is a possiblity of existence of an extra(, in 
addition, to a normal) set of 
pseudo-scalar and vector mesons.
Furthermore, the linear combinations of $C$ and $D$ are also applied, and normal and extra
sets of composite scalar and axial-vector mesons possibly exist as relativistic $S$-wave 
bound states. 

\begin{table}[t]
\caption{Level structure of the ground states of 
general quark meson systems:\ \ 
In $Q\bar Q$-meson system the approximate $LS$-symmetry is expected to be valid,
since $m_Q$ and $m_{\bar Q}>\Lambda_{\rm conf}$:
In $Q\bar q(q\bar Q)$-meson system the approximate chiral symmetry for light anti-quark
(quark) $\bar q(q)$ is to be valid, since $m_{\bar q}(m_q)\ll \Lambda_{\rm conf}$;
while the conventional heavy quark symmetry (HQS) is to be valid;  
In $q\bar q$ system the approximate chiral symmetry is 
to be valid.
 } 
\begin{center}
\begin{tabular}{|l|l|c|l|l|}
\hline\hline
Config. & Mass & Approx.Sym. & Spin WF& Meson Type\\
\hline
$Q\bar Q$ & $m_Q+m_{\bar Q}$ & $LS$ sym.
    & $u_Q(p_\mu )\bar v^{\bar Q}(p_\mu )
       =u({\mib p})\bar v({\mib p})$ & $P_s,V_\mu$ \\
\hline
$q\bar Q$ & $m_q+m_{\bar Q}$ & $q$:chiral sym.
     & $u_q(p_\mu )\bar v^{\bar Q}(p_\mu )
       =u({\mib p})\bar v({\mib p})$ & $P_s,V_\mu$ \\
          &                  & $\bar Q$:HQ Sym.
     & $u_q(-p_\mu )\bar v^{\bar Q}(p_\mu )
       =- v({\mib p})\bar v({\mib p})$ & $S, A_\mu$ \\
$Q\bar q$ & $m_Q+m_{\bar q}$ & $Q$:HQ Sym. 
     & $u_Q(p_\mu )\bar v^{\bar q}(p_\mu )
       =u({\mib p})\bar v^{\bar q}({\mib p})$ & $P_s,V_\mu$ \\
          &                  & $\bar q$:chiral sym.
     & $u_Q(p_\mu )\bar v^{\bar q}(-p_\mu )
       =u({\mib p})\bar u({\mib p})$ & $S, A_\mu$ \\
\hline
$q\bar q$ & $m_q+m_{\bar q}$ & chiral sym.
     & $\frac{1}{\sqrt 2}(u_q(p_\mu )\bar v^{\bar q}(p_\mu )
        \pm u_q(-p_\mu )\bar v^{\bar q}(-p_\mu ))$
     & $P_s^{(N)},V_\mu^{(N)}$; \\
          &      & 
     &  $\sim \frac{1}{\sqrt 2}(u({\mib p})\bar v({\mib p})
       \pm  v({\mib p})\bar u({\mib p}))$ & $P_s^{(E)},V_\mu^{(E)}$ \\
  & & & & \\
  &  &      & $\frac{(1,-i)}{\sqrt 2}(u_q(p_\mu )\bar v^{\bar q}(-p_\mu )
                \pm u_q(-p_\mu )\bar v^{\bar q}(p_\mu ))$
            & $S^{(N)},A_\mu^{(N)}$;  \\
  &  &  &  $\sim \frac{(1,-i)}{\sqrt 2}(u({\mib p})\bar u({\mib p})
                \pm v({\mib p})\bar v({\mib p}))$ & $S^{(E)},A_\mu^{(E)}$  \\
\hline\hline
\end{tabular}
\end{center}
\end{table}

In the above discussion 
we assume that $\Lambda_{\rm conf}\sim 1$GeV 
regardless of quark-flavor. 
The above expected level structure of ground states
is summarized in Table I.
 
\subsection{Level structure of excited state mesons}
In classifying the excited-state mesons we can proceed essentially
similarly as the case of ground state mesons. 
In the present approximation the masses of N-th excited states
are given by
$M_N^2=M_G^2+N\Omega\ (M_G\equiv M_0,\ \Omega$ being the inverse Regge slope), 
and their spin wave functions
are defined by the same formulas as given in \S 3 
with substitution of constituent 
exciton-quark mass $m_i$ by 
\begin{eqnarray}
m_i^* &=& \gamma_N\ m_i\ \ (\gamma_N\equiv M_N/M_G) . \label{eq51}
\end{eqnarray} 
The value of 
 $m_i^*$ for the ($q\bar q$) systems 
obtained by the formula (\ref{eq51}) 
are seen $m^* \ll \Lambda_{\rm conf}$ 
for the lower levels, especially for the ground and first-excited states,
and accordingly we can expect 
that chiral symmetry for these states may be still good. 
Similarly,  in the $(q\bar Q)/(Q\bar q)$ systems,
the chiral symmetry concerning the light quark 
is also good for the several lower levels.

\subsection{Expected spectroscopy of mesons}
In our fundamental equation Eq.(\ref{eq1})
the squared-mass spectra ${\cal M}^2$ is, as a first step  
in the pure-confining force limit, 
assumed to be Dirac-spinor independent.
Actually we must take into account the various effects due to one-gluon-exchange
potential and so on.\\
$\underline{light-quark\ (q\bar q)\ meson\ system}$:\ \ \  In the present
approximation
all the ground state mesons expected in \S 3 , $P_s^{(N,E)}, V_\mu^{(N,E)}, S^{(N,E)}$
and $A_\mu^{(N,E)}$ have the mass, $M_0=m_1+m_2$.
Actually the mass of $P_s^{(N)}$, assigned to $\pi$-nonet, 
should be exceptionally
low because of its nature\footnote{
The $q\bar q$-condensation are possible for the 
$S^{(N)}$, and $P_s^{(N)}$ other than $P_s^{(E)}$ is
a NG boson.
}   
as a Nambu-Goldstone boson. 
The masses of the $V_\mu^{(N)}$-nonet, assigned
to $\rho$-meson nonet, are known to be almost equal to
$M_0=m_1+m_2$. 
The masses of all the other ground-state mesons
are expected to be almost equal to those of the $V_\mu^{(N)}$-nonet,
and to be lower than those of the first-excited states.

For the first-excited states, the chiral symmetry is expected to be still 
effective, so we expect the
existence of a series of the first excited $P$-wave states of the ground state multiplets.
They have the masses, which are almost equal to the first excited states
of $V_\mu^{(N)}$-mesons and lower than the second excited states of them. 
Among the multiplets newly predicted in the present scheme, to be
called ``chiralons," the especially interesting mesons are the ones with 
the exotic quantum numbers; 
$J^{PC}$=$0^{+-}(S^{(E)}(S-wave)),\ 0^{--}(A_\mu^{(N)}(P-wave)),
\ 1^{-+}(S^{(E)}(P-wave) )$ and 
$1^{-+}(A_\mu^{(E)}(P-wave))$. 
Their masses, by the above mentioned estimate,
is expected to be, respectively, 
$
m(0^{+-})  \stackrel{<}{\scriptstyle \sim} 1.3\ {\rm GeV}$ and 
$
1.3\ {\rm GeV}  \simeq m(0^{--}) \simeq  m(1^{-+}) 
                 \simeq  m(1^{-+}) 
    \stackrel{<}{\scriptstyle \sim}  1.7\ {\rm GeV}.$\\
$\underline{heavy-light\ quark\ (Q\bar q\ and\ q\bar Q)\ meson\ system}$:\ \ \  As is seen 
from Table I we expect the existence of new $S$ and $A_\mu$ multiplets 
(at least the ground states).\\
$\underline{heavy\ quark\ (Q\bar Q)\ meson\ system}$:\ \ \  No new multiplets are expected
to exist. \\

\section{Experimental Evidences and Concluding Remarks}
In this paper we have presented a kinematical framework for describing 
covariantly the ground states as well as excited states of light-through-heavy
quark mesons. 

For light-quark mesons our scheme gives a theoretical basis
to classify the composite meson systems unifying the two contrasting viewpoints 
based on NRQM with $LS$ symmetry and on NJL model
with chiral symmetry. The essential physical assumption is to set up the 
bilocal Klein-Gordon equation (Yukawa equation) with
the squared-mass operator, which is, in the pure-confining force limit, independent
of Dirac-spinor suffix, and accordingly is chiral symmetric.
As a result is pointed out a possibility of existence of 
rather an abundant new nonets, chiralons, 
with masses lower than about 2 GeV; several ground 
and excited-state meson nonets.

For heavy/light quark meson systems 
we have similarly pointed out a possibility
of existence of new multiplets(triplets), chiralons.

Presently we can give a few experimental candidates for the predicted 
new multiplets: One of the most important and interesting ones is
the scalar $\sigma$ nonet (the members
are $\sigma (600)$,\cite{rfRR} $\kappa$(900),\cite{rfRR} 
$a_0(980)$ and $f_0(980)$) which 
constitutes,\cite{rfsca}
with $\pi$-nonet, a linear representation of the chiral $SU(3)$ symmetry.
It is notable that the $\sigma$ nonet is the relativistic $S$-wave state, 
which should 
be discriminated from the $^3P_0$ state.

Now the existence
of three pseudoscalars\cite{rf55} with mass between 1 GeV$\sim$1.5GeV  
($\eta$(1295), $\eta$(1420) and $\eta$(1460))
seems to be an established experimental fact. 
The two out of them may belong to the radially
excited $\pi$-nonet, while the one extra to the ground states of the
$P_s^{(E)}$-nonet newly predicted.

Also we have the other candidates for chiralons:
Recently it seems that the existence of two 
exotic particles\cite{rf6}
 in the $\eta\pi$ system with $J^{PC}=1^{-+}$ and 
with a mass around 1.5 GeV $\pi_1(1400)$  and 
$\pi_1(1600)$ has been accepted widely.
These two particles have the mass in the region estimated in \S 4.3
to be assigned as the respective excited $P$-wave states of $S^{(E)}$ and $A_\mu^{(E)}$.

We have the longstanding problem in hadron spectroscopy: the mass and width of
$a_1(1260)$ seem to be variant\cite{rf30} 
depending
 on the production process and/or decay channel.
We have made\cite{rfwaka} recently a preliminary 
analysis of the data obtained by GAMS group WA102 experiment on process
$\pi^- p\to 3\pi^0 n$, leading to an evidence of existence of two
$a_1(J^{PC}=1^{++})$ particles: $a_1^c(m=1.0{\rm GeV})$,
and $a_1^N(m=1.3{\rm GeV})$ (to be assigned, respectively, to 
$A_\mu^{(N)} (^3S_1)$, and to the conventional
$a_1$ particle).

Finally we should like to refer to a preliminary result of analysis on the 
heavy/light quark meson systems that a scalar chiralon
$B_0^\chi$ with $M\simeq 5.52$GeV may be observed\cite{rfR3}
in the $B\pi$ channel.


\begin{thebibliography}{9}
\bibitem{rf1}S. Ishida, M. Y. Ishida and M. Oda, 
   Prog. Theor. Phys. {\bf 93}, 939 (1995).
\bibitem{rf3}S. Ishida,
    Prog. Theor. Phys. {\bf 46}, 1570 and 1950 (1971). See also\\
R. P. Feynman, K. Kislinger and F. Ravndal, Phys. Rev. {\bf D3} (1971), 2706. 
\bibitem{rf2}H. Yukawa, Phys. Rev. {\bf 91}, 415 and 416 (1953). 
\bibitem{rfRR} K. Takamatsu; T. Tsuru; T. Komada; M. Ishida; in this workshop.
\bibitem{rfsca}
R. Delbourgo and M. D. Scadron, Phys. Rev. Lett. {\bf 48} (1982), 379.\\
M. Ishida, Prog. Theor. Phys. {\bf 101} (1999), 661.
\bibitem{rf55}S. Fukui et al., Phys. Lett. {\bf B267} (1991), 293.\\ 
D. Alde et al, Phys. Atomic Nuclei, {\bf 60} (1997), 386.\\
C. Caso et al., Euro. Phys. J. {\bf C3} (1998), 386, 398.\\
A. Ando et al., Phys. Rev. Lett. {\bf 57} (1986), 1296.\\
M. G. Rath et al., Phys. Rev. {\bf D40} (1989), 693.\\
Z. Bai et al., Phys. Rev. Lett. {\bf 65} (1990), 2507.
\bibitem{rf6}D. Alde et al., Phys. Lett. {\bf B205}(1988), 397.\\ 
H. Aoyagi et al.,  Phys. Lett. {\bf B314}(1993), 246.\\
S. U. Chung et al., Phys. Rev. {\bf D60}(1999), 092001;
in proc. of 8th International 
Conference on Hadron Spectroscopy (HADRON 99), Beijing, China, 24-28 Aug 1999.
\bibitem{rf30}S. Ishida, M. Oda, H. Sawazaki and K. Yamada,
Prog. Theor. Phys. {\bf 88} (1992), 89.
\bibitem{rfwaka}A. Wakabayashi, master thesis, Nihon University, 2000.
 M. Kobayashi, in this workshop.
\bibitem{rfR3} T. Maeda,  master thesis, Nihon University, 2000.
\end{thebibliography}
\end{document}